\documentclass[conference]{IEEEtran}

\IEEEoverridecommandlockouts

\def \shownames {Show names}
\def \showcomments {Show comments}

\ifdefined\showchanges

\else

\fi

\usepackage{algorithm,algorithmic}
\usepackage{amsmath}
\usepackage{amssymb}
\usepackage{balance}
\let\labelindent\relax
\usepackage{enumitem}
\usepackage[hang,flushmargin]{footmisc}
\usepackage[hyphens]{url}
\usepackage{hyperref}
\usepackage{letltxmacro}
\usepackage{listings}
\usepackage{msc}
\usepackage{tabularx}
\ifdefined\showcomments
\usepackage{todonotes}
\else
\usepackage[disable]{todonotes}
\fi

\LetLtxMacro{\todonote}{\todo}
\renewcommand{\todo}[2][]
{\todonote[inline, caption={#2}, size=\footnotesize, #1]
{\renewcommand{\baselinestretch}{0.5}\selectfont#2\par}}

\lstdefinelanguage{JavaScript}{
  keywords={typeof, new, true, false, catch, function, return, null, catch, switch, var, if, in, while, do, else, case, break},
  keywordstyle=\color{blue}\bfseries,
  ndkeywords={class, export, boolean, throw, implements, import, this},
  ndkeywordstyle=\color{darkgray}\bfseries,
  identifierstyle=\color{black},
  sensitive=false,
  comment=[l]{//},
  morecomment=[s]{/*}{*/},
  commentstyle=\color{purple}\ttfamily,
  stringstyle=\color{red}\ttfamily,
  morestring=[b]',
  morestring=[b]"
}

\usepackage{xspace}

\usepackage{balance}

\newcommand{\name}{\textsf{\small{S-FaaS}}\xspace}
\newcommand{\nameVarSize}{\textsf{S-FaaS}\xspace}

\newcommand{\tperiod}{\ensuremath{\tau}\xspace}
\newcommand{\tcur}{\ensuremath{t}\xspace}
\newcommand{\tmax}{\ensuremath{t_{max}}\xspace}
\newcommand{\deltat}{\ensuremath{\delta t}\xspace}
\newcommand{\mint}{\ensuremath{m_{int}}\xspace}
\newcommand{\mmax}{\ensuremath{m_{max}}\xspace}
\newcommand{\mavg}{\ensuremath{m_{avg}}\xspace}
\newcommand{\mem}{\ensuremath{m(t)}\xspace}
\newcommand{\tmem}{\ensuremath{t_{mem}}\xspace}
\newcommand{\deltam}{\ensuremath{\delta m}\xspace}
\newcommand{\netw}{\ensuremath{net}\xspace}

\begin{document}

\title{\nameVarSize: Trustworthy and Accountable Function-as-a-Service using Intel SGX}

\ifdefined\shownames

\author{
\IEEEauthorblockN{Fritz Alder\thanks{Author names listed in alphabetical order.}}
\IEEEauthorblockA{Aalto University\\
fritz.alder@acm.org}
\and
\IEEEauthorblockN{N. Asokan}
\IEEEauthorblockA{Aalto University\\
asokan@acm.org}
\and
\IEEEauthorblockN{Arseny Kurnikov}
\IEEEauthorblockA{Aalto University\\
arseny.kurnikov@aalto.fi}
\and
\IEEEauthorblockN{Andrew Paverd}
\IEEEauthorblockA{Aalto University\\
andrew.paverd@ieee.org}
\and
\IEEEauthorblockN{Michael Steiner}
\IEEEauthorblockA{Intel Labs\\
michael.steiner@intel.com}
}

\fi

\maketitle

\begin{abstract}
Function-as-a-Service (FaaS) is a recent and already very popular paradigm in cloud computing.
The function provider need only specify the function to be run, usually in a high-level language like JavaScript, and the service provider orchestrates all the necessary infrastructure and software stacks.
The function provider is only billed for the actual computational resources used by the function while it is running.
Compared to previous cloud paradigms, FaaS requires significantly more fine-grained resource measurement mechanisms, for example to measure the compute time and memory usage of a single function invocation with sub-second accuracy.
Thanks to the short duration and stateless nature of functions, and the availability of multiple open-source frameworks, FaaS enables small ephemeral entities (e.g.\ individuals or data centers with spare capacity) to become service providers.
However, this exacerbates the already substantial challenge of ensuring the resource consumption of the function is measured accurately and reported reliably.
It also raises the issues of ensuring the computation is done correctly and minimizing the amount of information leaked to the service provider.

To address these challenges, we introduce \name{}, the first architecture and implementation of FaaS to provide strong security and accountability guarantees backed by Intel SGX.
To match the dynamic event-driven nature of FaaS, our design introduces a new key distribution enclave and a novel \emph{transitive attestation} protocol.
A core contribution of \name{} is our set of resource measurement mechanisms that securely measure compute time inside an enclave, and actual memory allocations.
We have integrated \name{} into the popular OpenWhisk FaaS framework.
We evaluate the security of our architecture, the accuracy of our resource measurement mechanisms, and the performance of our implementation, showing that our resource measurement mechanisms add less than 6.3\% performance overhead on standardized benchmarks.
\name{} can be integrated with smart contracts to enable decentralized payment for outsourced computation.

\end{abstract}

\section{Introduction}
\label{sec:introduction}

Function-as-a-Service (FaaS) is a recent paradigm in outsourced computation that has generated significant interest from cloud providers and developers.
The function provider need only specify the function to be performed, whilst the service provider provides all the infrastructure necessary to run, scale, and load-balance the function.
Compared to the Infrastructure-as-a-Service (IaaS) paradigm, FaaS significantly simplifies the task of the function provider, who no longer needs to requisition a specific number of virtual machines (VMs), or install and maintain a full software stack.
FaaS also improves efficiency for the service provider, who can now optimize the underlying infrastructure to maximize performance and throughput, whilst only isolating individual functions from one another.

The function provider only pays for the computational resources actually used by the function, instead of being billed to run a number of full-stack VMs.
For example, Amazon Web Service (AWS) Lambda is billed based on the number of function invocations and the time-integral of the function's memory usage, measured in Gigabyte-seconds (GB-s)~\cite{aws_lambda_pricing}.
Google Cloud Functions follows a similar billing structure, but also includes raw compute time, measured in Gigahertz-seconds (GHz-s), and network usage, measured in GB~\cite{google_cloud_functions_pricing}.

In addition to large cloud providers, the highly dynamic nature of FaaS also makes it possible for smaller entities to become \emph{ephemeral} FaaS service providers. 
Specifically, compared to the long-lived VMs in the IaaS paradigm, the short runtimes and stateless nature of FaaS functions make it possible for data centres with spare capacity or even individuals with suitably powerful PCs and stable internet connections (e.g.\ gamers) to become FaaS service providers. 
The idea of outsourcing computation to individuals or small service providers has existed for many years, and has spawned successful projects like SETI@home~\cite{setiathome}, Folding@home~\cite{foldingathome}, and ClimatePrediction.net~\cite{climateprediction}.
However, these perform fixed computations, and on a voluntary basis.
A new trend, evidenced by recent projects like the Golem network~\cite{golem_network}, aims to support arbitrary computation and also to remunerate the entities who perform the computation.

Any type of outsourced computing raises multiple security concerns, but FaaS arguably makes these more acute.
Firstly in terms of \emph{integrity}, the function provider has no control of the underlying software stack, and so has less certainty that the function is being executed correctly.
Secondly in terms of \emph{confidentiality}, the functions' inputs and outputs are directly visible to the service provider.
Finally, in terms of \emph{accountability}, the fine-grained sub-second resource measurements in FaaS are significantly more difficult to audit than the coarse-grained billing of IaaS, where usage is typically measured per VM per hour~\cite{aws_ec2_spot_pricing}.
At present, function providers are required to trust that service providers are honest, based solely on the latter's reputation.
Although this may be acceptable in the case of large cloud providers, it would almost certainly preclude small ephemeral service providers from providing FaaS services.
The challenge is therefore to provide security and accountability for FaaS services hosted by \emph{any} service provider.

To overcome these challenges, we present \name{}, an approach for providing security and accountability in FaaS using Intel Software Guard Extensions (SGX).
SGX is a modern Trusted Execution Environment (TEE) providing hardware-based isolation and protection for code and data inside an \emph{enclave}. 
Our \name{} architecture encapsulates individual functions inside SGX enclaves, and uses remote attestation to provide various guarantees to relying parties.
Specifically, \name{} enhances \emph{security} by protecting the integrity and authenticity of the function inputs and outputs, and provides strong assurance to the client that the outputs are the result of a correct execution of the function with the given inputs.
\name{} provides \emph{accountability} by producing a verifiable measurement of the fine-grained resource usage of each function invocation.
Crucially, this measurement can be verified by both the service provider and the function provider, even though the latter does not control the software stack.
\name{} can also enhance \emph{privacy} by hiding the function's inputs and outputs from the service provider.
However, given the various side-channel attacks that have recently been demonstrated against SGX, we only claim to enhance privacy for a specific class of functions in which the control flow and memory access patterns are input-independent.
Defending against such side-channel attacks is an orthogonal challenge to our work.

To the best of our knowledge we are the first to investigate trustworthiness and metering of outsourced computation in the context of FaaS.
We provide a full implementation and evaluation of our architecture in order to 1) investigate the subtleties of such a design (e.g.\ in terms of integrating it into existing frameworks), and 2) to measure the performance overhead of this type of architecture.
The ability to accurately measure resource usage of SGX enclaves is a new and challenging problem, since it goes well beyond the current functionality provided by SGX.
Inspired by the \emph{reference clock} developed by Chen et al.~\cite{chen_detecting_2017}, we develop a set of accurate and trustworthy resource measurement mechanisms for SGX enclaves, using Intel Transactional Synchronization Extensions (TSX), which can be deployed without any hardware changes.
Although we demonstrate them in the context of FaaS, these mechanisms can be used in various other SGX applications.
We have integrated \name{} into the Apache OpenWhisk FaaS framework~\cite{openwhisk}.

In summary, we claim the following contributions:

\begin{itemize}

\item \textbf{Design of \name{}:} We develop an architecture for protecting FaaS deployments using Intel SGX that 1) ensures the integrity (and in some cases confidentiality) of function inputs and outputs, and 2) provides clients with strong assurance that the outputs are the result of a correct execution of the function with the given inputs (Section~\ref{sec:architecture}).

\item \textbf{SGX resource measurement:} We present a set of mechanisms for accurately measuring the compute time, memory, and network usage of a function executing inside an SGX enclave, in a manner that can be trusted by \emph{both} the service provider and function provider (Section~\ref{sec:resource_measurement}).

\item \textbf{Full implementation of \name{}:} We provide a proof-of-concept implementation of the architecture and resource measurement mechanisms as part of the OpenWhisk FaaS framework (Section~\ref{sec:openwhisk}).
We systematically evaluate \name{} and show that its resource measurement mechanisms are accurate and that it can provide all security guarantees with minimal performance overhead (Section~\ref{sec:evaluation}).

\end{itemize}

\section{Background}

\subsection{Function as a Service}
\label{sec:faas}

\begin{figure}
\includegraphics[width=\columnwidth]{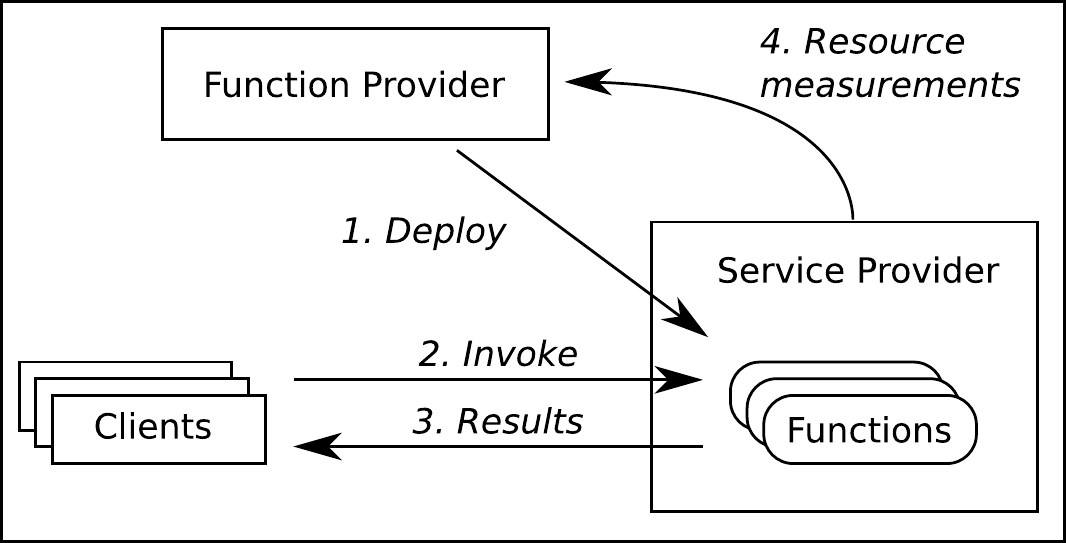}
\caption{Main entities and interactions in FaaS.}
\label{fig:system_model}
\end{figure}

Figure~\ref{fig:system_model} shows a generalized overview of a FaaS scenario.
The computational infrastructure on which the function is executed is operated by the \emph{service provider}.
We distinguish between the entity that provisions the function, i.e.\ the \emph{function provider}, and the entity that invokes the function, i.e.\ the \emph{client}.
The client supplies the inputs on which the function is run, and receives the corresponding output.
The service provider bills the function provider for the resources used by the function.
In some cases, the function provider could be the sole client.
For example, an enterprise could outsource an infrequently used but computationally intensive function (e.g.\ generating reports from historical data) to a FaaS service, and only allow it to be invoked by authorized employees.
Alternatively, the function provider can determine which other entities may access the function as clients.
For example, to deal with highly variable demand, a web developer could provision a function that can be directly invoked from authorized users' web browsers (possibly via an API gateway like that provided by AWS~\cite{aws_api_gateway}).

In IaaS, billing policies typically consider the number of hours for which a VM is running.
In contrast, a single FaaS function may run for less than a second, and thus FaaS requires very fine-grained resource measurements.
The following types of resource measurements are used in the billing policies of different FaaS service providers:

\begin{itemize}

\item \textbf{Function invocations:} the number of times the function is called.

\item \textbf{Compute time:} the execution time of the function multiplied by the frequency of the CPU, typically measured in GHz-seconds (GHz-s), which is dimensionally equivalent to CPU cycles.

\item \textbf{Memory time-integral}: the execution time of the function multiplied by the amount of memory provisioned or used, measured in Gigabyte-seconds (GB-s). 
This can be calculated using either the maximum allowable memory, the maximum allocated memory, the average allocated memory, or the time-varying memory allocation. 

\item \textbf{Network usage:} the amount of data sent or received, measured in Gigabytes (GB).

\end{itemize}

Table~\ref{tab:faas_examples} gives examples of different FaaS service providers and the types of resource measurements used in their billing policies.
Note that calculating the memory time-integral inherently requires measuring the execution time of the function (indicated by $\circ$), even if the compute time itself is not used directly in the billing policy.

\begin{table}[H]
\renewcommand*{\arraystretch}{1.3}
\caption{Billing policies of some current FaaS services}
\label{tab:faas_examples}
\begin{tabularx}{\columnwidth}{|X|c|c|c|c|}
\hline
\textbf{Service}       & \textbf{Invocations} & \textbf{Time} & \textbf{Memory} & \textbf{Network}\\
\hline
AWS Lambda~\cite{aws_lambda_pricing}                          & \checkmark & $\circ$    & \checkmark &            \\
\hline
Azure Functions~\cite{azure_functions_pricing}                & \checkmark & $\circ$    & \checkmark &            \\
\hline
Google Cloud Functions~\cite{google_cloud_functions_pricing}  & \checkmark & \checkmark & \checkmark & \checkmark \\
\hline
IBM Cloud functions~\cite{ibm_cloud_functions_pricing}        & \checkmark & $\circ$    & \checkmark &            \\
\hline
\end{tabularx}
\end{table}

\subsection{Intel SGX}
\label{sec:sgx}

Intel Software Guard Extensions (SGX) technology is a set of CPU extensions that allow applications to instantiate isolated execution environments, called \emph{enclaves}, containing application-defined code.
An enclave runs as part of an application process (i.e.\ the \emph{host application}) in the applications virtual address space, but data inside an enclave can only be accessed by code within the same enclave.
Code within the enclave can only be called from the untrusted host application via well-defined call gates.
An ECALL refers to untrusted code calling a function inside the enclave, thus transferring control to the enclave, and an OCALL refers to enclave code calling an untrusted function outside the enclave.
SGX therefore protects the integrity of enclave code and the confidentiality and integrity of enclave data against all other software on the platform, including privileged system software like the OS and hypervisor.

By design, enclave code can be interrupted at any time by the untrusted software (e.g.\ to allow the OS to schedule another process).
This is referred to as an Asynchronous Enclave Exit (AEX).
After the interrupt has been completed, the host application can resume the enclave's operation using the ERESUME instruction, which continues the enclave's execution from the point at which it was interrupted.
Critically, this AEX and ERESUME are handled by the the CPU, and are thus transparent to the enclave's code.

As part of its internal data structures, an enclave allocates one or more Thread Control Structure (TCS) data structures in protected memory.
The number of TCS data structures determines the maximum number of threads that may enter the enclave concurrently.
Whenever a thread enters the enclave, it is associated with a free TCS, which it marks as \texttt{busy} and uses to store state information.
Specifically, the enclave allocates a stack of Save State Areas (SSAs) for each TCS, and the TCS contains a pointer to the current SSA (CSSA).
When enclave code is interrupted, the CPU stores its current register values in the current SSA for that thread, and fills the registers with synthetic contents before switching to the untrusted code, to avoid leaking secrets.
When an enclave thread is resumed (via ERESUME), the CPU reloads its registers from the SSA and continues executing. 
Code inside the enclave cannot read the TCS data structures, but can read and modify SSAs.

Every enclave has an \emph{enclave identity}, called the MRENCLAVE value, which is a cryptographic hash of the enclave's configuration, e.g.\ memory pages, at the time it was initialized.
Enclaves containing precisely the same code and configuration will have the same MRENCLAVE value, even if they are run on different physical hardware platforms.
An enclave has a \emph{signing identity}, called the MRSIGNER value, which is the hash of the public key of the developer who signed the enclave.

In addition to isolated execution, SGX provides \emph{sealed storage} by allowing each enclave to encrypt (i.e.\ \emph{seal}) persistent data so that it can be stored outside the enclave.
Data can be sealed either against MRENCLAVE, such that it can only be unsealed by precisely the same enclave running on the same physical platform.
Alternatively, data can be sealed against MRSIGNER, such that it can be unsealed on the same platform by any enclave signed by the same developer key.

SGX also provides \emph{remote attestation}, a process through which an enclave can prove its identity (i.e.\ its MRENCLAVE and MRSIGNER values) to a remote verifier.
Specifically, SGX creates a \emph{quote} consisting of the enclave's identity and a small amount of user-defined data (e.g.\ hashes of public keys generated by the enclave).
This quote can be verified using the Intel Attestation Service (IAS).

\subsection{Intel TSX}
\label{sec:tsx}

Intel Transactional Synchronization Extensions (TSX)\footnote{\url{http://www.intel.com/software/tsx}} is an instruction set extension, available since the Haswell microarchitecture, providing transactional memory support in hardware.
TSX is designed to improve performance of concurrent programs by reducing the use of software-based lock primitives (e.g.\ mutex or spinlock).
TSX can be used inside an SGX enclave.

We focus on four of the new instructions introduced by TSX, namely \texttt{XBEGIN}, \texttt{XEND}, \texttt{XABORT}, and \texttt{XTEST}.
The \texttt{XBEGIN} and \texttt{XEND} instructions are used to designate the beginning and end of a transactional region of code.
The \texttt{XABORT} instruction aborts a currently executing transaction.
\texttt{XTEST} is used to test if the thread is currently within a transaction.

For each transaction, the CPU maintains a read-set and write-set, consisting of all data memory addresses (at cache-line granularity) that will be read or written by the transaction.
The CPU monitors accesses to these addresses by other threads, and if a conflicting access is detected, the transaction is aborted.
In particular, the transaction will be aborted if any other thread writes to an address in the transaction's read-set.
A transaction is also aborted by any asynchronous exception (e.g.\ an OS interrupt).

If a transaction is aborted, the CPU will not commit any writes from the transactional region to memory.
The CPU will execute the transaction's abort handler, which was registered when the transaction was started.
This handler can decide whether to re-attempt the hardware transaction, or proceed to a fallback path, which typically uses a software lock primitive.

\section{Threat Model and Requirements}

\subsection{Threat Model}
\label{sec:threat_model}

Given the context in which a FaaS deployment operates, we assume the following two types of adversaries:

The \textbf{service provider} could be adversarial from the perspective of the function provider and the clients.
This entity has the capability to run arbitrary software on the server platforms, including privileged software and SGX enclaves.
Specifically, with full control of the OS, this entity can interrupt and resume enclaves at any time e.g.\ using the SGX-Step framework~\cite{van_bulck_sgx-step:_2017} to interrupt the enclave with single-instruction granularity.
An adversarial service provider could have the following objectives:

\begin{itemize}

\item Learn the inputs and outputs of the specific function invocations.

\item Modify the inputs and outputs, or execute the function incorrectly.

\item Overcharge the function provider, either by falsely inflating resource usage measurements or making fake requests to the function.

\end{itemize}

The \textbf{function provider} could be adversarial from the perspective of the service provider, and has the ability to submit arbitrary functions to the service.
We assume that the function provider is not adversarial from the perspective of the clients, since they have decided to use the function.
This adversary could have the following objective:

\begin{itemize}

\item Under-report resources used by the function.

\end{itemize}

We assume it is infeasible for any entity to subvert correctly-implemented cryptographic primitives.
Although various side-channel attacks have been demonstrated against SGX (e.g.~\cite{xu_controlled-channel_2015,bulck_telling_2017,lee_inferring_2017,brasser_software_2017,chen_sgxpectre_2018}), defending against these is an orthogonal challenge, as we discuss in Section~\ref{sec:discussion}.

\subsection{Requirements}
\label{sec:requirements}

Based on the above adversary capabilities and objectives, we define the following requirements:

\begin{itemize}

\item[\textbf{R1}] \textbf{Security:} The service provider must be prevented from modifying the inputs or outputs of a function invocation, and the client must receive assurance that output $\mathcal{O}$ is the result of a correct execution of the intended function $\mathcal{F}$ on the supplied inputs $\mathcal{I}$.

\item[\textbf{R2}] \textbf{Privacy:} For functions exhibiting input-independent control flow and memory access patterns, the service provider must be prevented from learning the inputs $\mathcal{I}$ or outputs $\mathcal{O}$ of a function invocation.

\item[\textbf{R3}] \textbf{Measurement accuracy:} The system must produce an accurate measurement of the resource usage of each function invocation, consisting of the total time for which the function was executing, the time-integral of the function's memory usage, and the total network traffic sent and received by the function.

\item[\textbf{R4}] \textbf{Measurement veracity:} Both the service provider and function provider must be able to verify the authenticity of the resource measurement. 

\end{itemize}

In some cases, it may also be desirable to hide the behavior of the function itself from the service provider (e.g.\ a \emph{secret} algorithm).
Although this may be possible in certain cases, the ability to hide the code executing in the enclave is not part of the security guarantees of SGX, and so is beyond the scope of our current work.

\section{Design Challenges}
\label{sec:challenges}

Securing a FaaS system using SGX requires solving multiple challenges.
In this section we outline the principal challenges, and the associated sub-challenges arising from these.

\subsection{Dynamic server utilization}
\label{sec:challenges_dynamic}

Even compared to existing cloud computing paradigms like IaaS, FaaS is significantly more dynamic in terms of server utilization.
For example, the popular OpenWhisk FaaS framework spawns a new worker environment for \emph{each invocation} of a function.
These worker environments can be spawned on any available physical server.
Furthermore, a worker environment may only be provisioned with the function it will run \emph{after} the system receives an invocation request for that function from the client.
Whilst this gives service providers significant flexibility (e.g.\ to perform load balancing), it poses multiple challenges if the function is to be run within an SGX enclave.

\textbf{C1 Encrypting client input:}
The first challenge is that, to achieve Requirements \textbf{R1} and \textbf{R2}, the client's inputs must be encrypted \emph{before} knowing which worker enclave will run the function.
We therefore require a type of transferable state, consisting of a set of cryptographic keys, that can be distributed to any worker enclave.
Since the service provider is potentially adversarial, we cannot trust this entity to store and distribute the state.
To solve this, our architecture includes a new \emph{key distribution enclave (KDE)} to securely generate and distribute the necessary keys to the dynamically created worker enclaves (Section~\ref{sec:key_distribution}).

\textbf{C2 Attesting worker enclaves:}
The second challenge is that this dynamic server utilization and one-shot function invocation precludes the use of standard SGX remote attestation, in which a client would establish the trustworthiness of a specific worker enclave via a multi-round-trip protocol, before sending any secrets to the enclave.
Even if the client's input is encrypted using keys from the KDE, the client still needs to establish the same level of trust as if it had attested the worker enclave.
We solve this by developing a new type of \emph{transitive attestation} scheme (Section~\ref{sec:transitive_attestation}).

\subsection{Fine-grained resource usage measurement}
\label{sec:challenges_resources}

As explained in Section~\ref{sec:faas}, FaaS requires an accurate mechanism for measuring the compute time of a function, either to be used directly in the billing policy, or to calculate other billing metrics like the time-integral of the function's memory usage.
Even if a function is run inside an SGX enclave, we still require an accurate and trustworthy mechanism to measure the enclave's compute time, to achieve requirements \textbf{R3} and \textbf{R4}.

\textbf{C3 Measuring time in enclaves:}
SGX does not include any direct functionality to securely measure execution time of an enclave. 
In particular, although the \texttt{rdtsc} instruction ostensibly returns the number of cycles since reset, this value can be modified by a malicious OS or hypervisor.
The \texttt{sgx\_get\_trusted\_time} function included in the SGX SDK cannot be used for measuring an enclave's execution time because it can be arbitrarily delayed by the OS, since it requires the enclave to make an OCALL.
Furthermore, any timing mechanism must account for the fact that the OS can interrupt the enclave at any point in its execution (causing an AEX), wait for an arbitrary period of time, and then transparently resume the enclave using ERESUME.
To solve this challenge, we develop a custom mechanism for measuring the time spent executing a function inside the enclave (Section~\ref{sec:timing}), using Intel TSX.

\textbf{C4 Sandboxing untrusted functions:}
One constraint of our timing mechanism is that it must be run in the same enclave as the function it measures, which gives rise to a second order challenge.
To fulfil requirement \textbf{R4}, \emph{both} the service provider and function provider must be able to trust the timing measurement, which is challenging given that we assume each may be adversarial from the other's perspective (Section~\ref{sec:threat_model}).
From the function provider's perspective, the enclave protects the measurement from manipulation by the service provider.
However, from the service provider's perspective, the enclave does not protect the measurement against malicious functions supplied by the function provider.
To address this, we ensure that the function is \emph{sandboxed} within the enclave, in order to protect the resource measurement mechanisms (Section~\ref{sec:architecture}).

\section{Architectural Design}
\label{sec:architecture}

\begin{figure}
\includegraphics[width=\columnwidth]{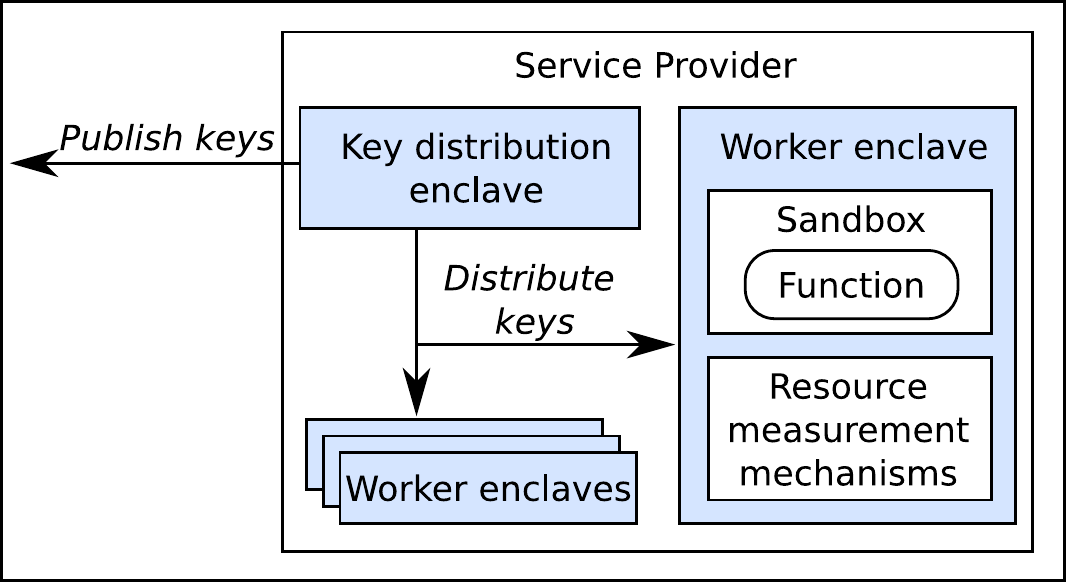}
\caption{Service Provider architecture in \nameVarSize{}.}
\label{fig:architecture}
\end{figure}

In this section we present our overall \name{} architecture.
We first describe the principal entities and then discuss the main types of operations and interactions between these entities.

Figure~\ref{fig:architecture} shows an overview of the service provider in our architecture.
The service provider runs a centralized \emph{key distribution enclave (KDE)} and a variable number of \emph{worker enclaves}.
The worker enclaves can be run on different physical servers.

\textbf{Key distribution enclave (KDE):}
To address challenges \textbf{C1} and \textbf{C2} described in Section~\ref{sec:challenges_dynamic}, we introduce a KDE that is responsible for securely generating and distributing keys to various entities.
The KDE pre-generates the necessary key pairs and then distributes the public keys to clients and the corresponding private keys to worker enclaves.
For scalability reasons, a service provider may run multiple KDEs.
Clients can be directed to any KDE, but client requests encrypted under keys from a specific KDE can only be processed by worker enclaves that have also obtained keys from the same KDE.

\textbf{Worker enclave:}
Worker enclaves are responsible for running the functions with the provided inputs and measuring the functions' resource usage.
As shown in Figure~\ref{fig:architecture}, each worker enclave consists of two main subsystems: a sandboxed interpreter, in which the function is run, and our resource measurement mechanisms.
This design is motivated by the following two factors:
First, for performance reasons, the service provider will typically create a pool of worker enclaves and then dynamically provision functions to worker enclaves as requests are received.
Our use of an interpreter allows any worker enclave to run any function.
This also has the benefit that all worker enclaves share the same enclave identity (MRENCLAVE value), which simplifies key distribution and attestation.
Second, as outlined in Section~\ref{sec:challenges_resources}, one important constraint of our resource measurement mechanisms is that they must be run within the same enclave as the function they measure, and must thus be protected from other code running in the same enclave (Challenge \textbf{C4}).
By running the function in a sandboxed interpreter, our design inherently solves this challenge.
Even without our architecture, a FaaS service provider would always run functions in some type of sandbox in order to isolate different functions from one another and protect the underlying infrastructure from potentially malicious functions.
If using \name{}, the service provider does not need an additional sandbox.  
The specific operations performed by our worker enclave are described in Section~\ref{sec:worker_enclave}

\begin{figure*}
\includegraphics[width=\textwidth]{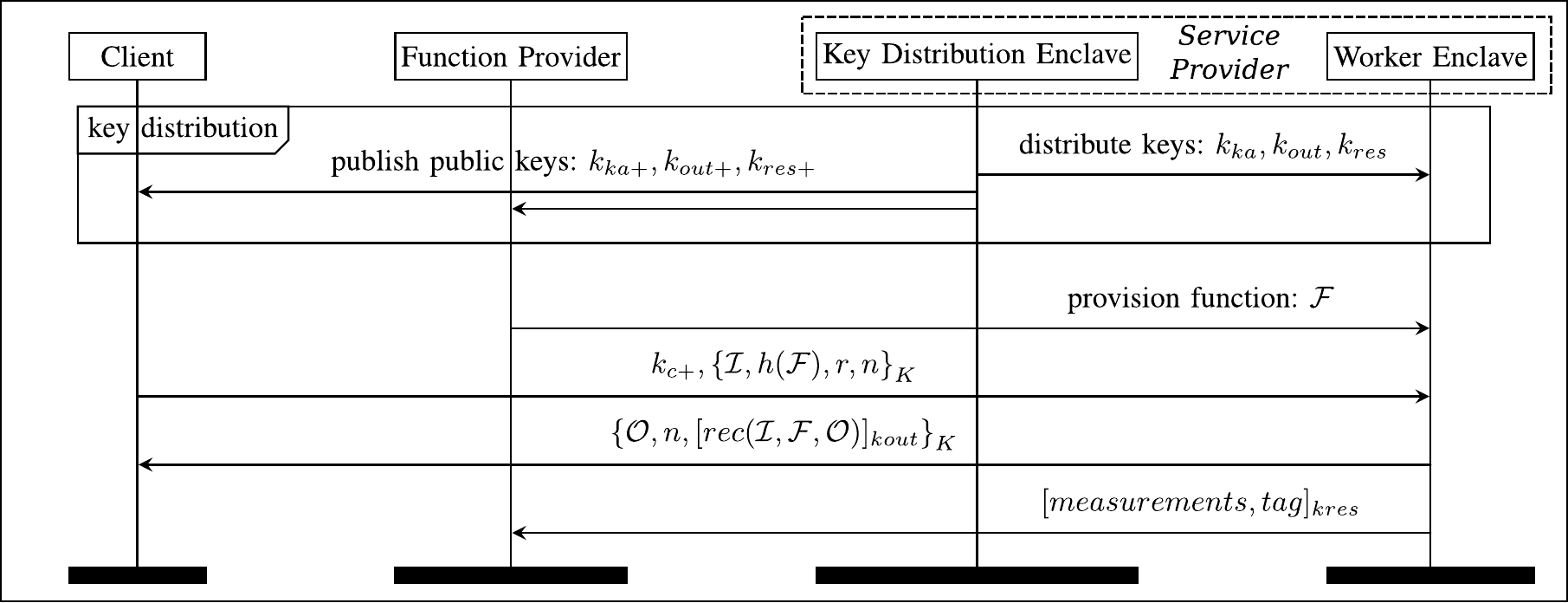}
\caption{Sequence of interactions between the provisioning enclave, worker enclave, function provider, and client. $k_{x+}$ denotes the public key corresponding to $k_{x}$, $h()$ denotes a cryptographic hash function, $\{\}_K$ denotes an encryption under key $K$, and $[]_S$ denotes a signature using key $S$.}
\label{fig:interactions}
\end{figure*}

Figure~\ref{fig:interactions} shows a high-level overview of the sequence of interactions between the function provider, KDE, worker enclave, and client.
We explain these interactions in detail in the following sections.

\subsection{Key Distribution}
\label{sec:key_distribution}

The KDE pre-generates a \emph{key set}, consisting of the following three asymmetric key pairs:

\begin{itemize}

\item A \textbf{key agreement key $k_{ka}$}, which is used to authenticate the worker enclave to clients and establish a shared session key $K$ with each client.

\item An \textbf{output signing key $k_{out}$}, which is used to create a type of signed \emph{receipt} indicating that the outputs are the result of a correct invocation of the executed function for the given inputs.
This receipt can be verified by entities other than the client, and is only produced if requested by the client.

\item A \textbf{resource measurement signing key $k_{res}$}, which is used to sign the final resource measurement for each invocation of the function.

\end{itemize}

As shown in the first part of Figure~\ref{fig:interactions}, the KDE distributes these keys to the worker enclaves.
Depending on the service provider's configuration, the same key set can be provisioned to any number of worker enclaves.
Sharing a key set amongst more worker enclaves makes load balancing easier, but increases the risk of having to revoke the key set if any worker enclave is physically compromised.
The service provider can also rotate the key sets as frequently as required.

The KDE also distributes the corresponding public keys to the function provider and clients.
A client can thus use the public key agreement key $k_{ka+}$ in conjunction with her own key agreement key $k_c$ to generate a symmetric session key $K$.
For example, this could be implemented using elliptic curve Diffie-Hellman key agreement.
When communicating with the enclave, the client will send her public key $k_c+$, and then encrypt all further information under the session key $K$, thus solving Challenge \textbf{C1}.

\subsection{Transitive Attestation}
\label{sec:transitive_attestation}

As explained in Challenge \textbf{C2} (Section~\ref{sec:challenges_dynamic}), in a typical FaaS deployment it is not possible to have each client and function provider (i.e.\ the \emph{relying parties}) perform a remote attestation of the worker enclave for each function invocation.
Instead, we use a new type of transitive attestation approach to allow the relying parties to establish an equivalent level of trust as if they had attested the relevant worker enclave, whilst only attesting the centralized KDE.

When the KDE generates a new key set, it produces an SGX quote that can be verified using the Intel Attestation Service (IAS).
This quote includes the enclave identity of the KDE ($\mathcal{MRE}_{KDE}$) as well as the hashes of the three generated public keys ($k_{ka+}$, $k_{out+}$, and $k_{res+}$) and the enclave identity of the worker enclaves ($\mathcal{MRE}_{WE}$).
By design, the KDE will only distribute the corresponding private keys to a worker enclave whose identity matches $\mathcal{MRE}_{WE}$.

When a worker enclave requires its key set, the KDE attests the worker enclave, and checks that the attested identity matches $\mathcal{MRE}_{WE}$.
The service provider can also implement additional policies outside the KDE, e.g.\ to ensure that keys can only be distributed to physical machines within the service provider's data center.

Therefore, the full chain of trust is as follows:
\begin{enumerate}
\item The relying parties attest the KDE to verify that i) a particular key set was generated by a valid KDE, and ii) the KDE will only distribute those keys to a specific type of worker enclave;
\item The KDE attests each worker enclave and checks its identity before distributing the key set;
\item The worker enclave interacts with the relying parties using the key set it received, thus establishing a transitive trust relationship between the relying parties and worker enclave.
\end{enumerate}

\subsection{Function Provisioning \& Measurement}

The function provider is assumed to have a business relationship with the service provider, including a means to authenticate itself (e.g.\ username and password, or public key).
The service provider uses this mechanism to control who may provision functions.
These intermediate steps are not shown in Figure~\ref{fig:interactions}, but the overall result is that a function $\mathcal{F}$, provided by the function provider, is eventually provisioned to a worker enclave.

As the function runs, the resource measurement mechanisms in the worker enclave monitor the function's resource usage.
Once the function completes, these mechanisms finalize the measurements and output a data structure containing the metrics shown in Table~\ref{tab:resources}.
Section~\ref{sec:resource_measurement} explains how we measure each of these quantities.

\begin{table}[H]
\renewcommand*{\arraystretch}{1.3}
\caption{Resource Measurement Metrics}
\label{tab:resources}
\begin{tabularx}{\columnwidth}{|l|X|}
\hline
\tmax & Total compute time of the function, reported as a multiple of \tperiod\\
\hline
\tperiod & Duration of each tick in CPU cycles\\
\hline
\mint & Time-integral of memory usage\\
\hline
\mmax & Maximum memory used by the function\\
\hline
\mavg & Average memory used by the function\\
\hline
\netw & Total number of network bytes sent and received\\
\hline
\end{tabularx}
\end{table}

In addition, the function itself outputs a fixed-size $tag$ to be included in the resource measurement data structure.
This can be used by the function provider to check that each resource measurement corresponds to a valid invocation of the function, thus preventing a malicious service provider from spuriously running the function to drive up resource usage. 
For example, this tag could be the cryptographic hash of the client's API key or authentication token, which the client provided as part of the encrypted input.

Finally, the worker enclave signs a hash of the complete resource measurement data structure using $k_{res}$.
This signed measurement is eventually returned to the function provider, either immediately or at the end of a billing period.

\subsection{Function Invocation \& Output}

Once a client has obtained the published set of public keys from the KDE (Section~\ref{sec:key_distribution}), and verified the KDE's attestation (Section~\ref{sec:transitive_attestation}), she runs the key agreement protocol to generate a symmetric session key $K$ under which to encrypt her inputs using an authenticated encryption algorithm (e.g.\ AES-GCM).
Note that this key exchange only achieves unilateral authentication of the enclave towards the client, since the enclave's public key is authenticated via the transitive attestation.
We assume that any client authentication, if required, is included in the inputs supplied by the client (e.g.\ an API key or authentication token).

In addition to encrypting the inputs for the worker enclave, the client also wants to ensure that only the intended function may process her inputs.
If this check were omitted, a trivial attack would be to redirect the client's input to a function that simply publishes any input it receives.
Unlike typical systems using remote attestation, this check is necessary in our case because the function is dynamically provisioned to the worker enclave \emph{after} the worker enclave has been attested by the KDE.
To achieve this, the client includes a hash of the intended function $h(\mathcal{F})$ as part of her encrypted input.
Upon receiving this, the worker enclave checks that this matches the hash of the function it has loaded, and if not, aborts the invocation \emph{before} passing the decrypted inputs to the function.

The client includes a flag $r$ to indicate whether the worker enclave should produce a \emph{receipt} of the invocation, and a nonce $n$ to associate a specific request message with a response.
If the client requests a receipt, the enclave produces a receipt data structure $rec(\mathcal{I}, \mathcal{F}, \mathcal{O})$, signed using $k_{out}$, certifying that the output $\mathcal{O}$ is the result of executing function $\mathcal{F}$ on input $\mathcal{I}$.
To facilitate billing policies in which the client pays for the computation, the client can also request to receive the resource measurement data structure described above, and have a hash of this included in the receipt.

Once the function completes, the worker enclave returns the outputs and the nonce to the client via the encrypted channel, and includes the signed resource measurement and receipt, if requested.
If the function terminates abnormally, the worker enclave sends an error message to the client in lieu of the output.

\subsection{Worker Enclave Operations}
\label{sec:worker_enclave}

The worker enclave performs four main operations, each of which is initiated by an ECALL, as shown in Listing~\ref{lst:worker_ecalls}.

\begin{lstlisting}[frame=single, language=C, caption={Worker Enclave ECALLs}, captionpos=b, float, frame=tblr, label=lst:worker_ecalls]
uint32_t ecall_setup(key_dist_enclave_addr, 
    sealed_data*)

uint32_t ecall_init(function*, function_size)

uint32_t ecall_run(encrypted_input*, output_size*)

uint32_t ecall_finish(encrypted_output*, measurements*, 
    measurement_signature*)

\end{lstlisting}

\textbf{\texttt{setup:}} This ECALL triggers the worker enclave to either obtain its key set from the KDE, or unseal a previously-received key set.
If a new key set is obtained, the enclave will seal this against its own MRENCLAVE value, and return the sealed data to the host application.
Note that this sealed data can only be unsealed by the same enclave running on the same physical server.
When the service provider rotates the key set, this ECALL is used to trigger the worker enclave to obtain the new key set from the KDE.
This ECALL is performed as a preparatory step before any function is provisioned, and thus does not contribute any performance overhead to a function invocation.

\textbf{\texttt{init:}} When a function is provisioned, this ECALL takes the specified function as a parameter and loads this into the sandbox.
The service provider can configure whether the time required to load the function is included in the total compute time measurement.

\textbf{\texttt{run:}} When a function invocation is requested, this ECALL is called with the encrypted inputs as parameters.
The enclave first calculates the shared session key $K$ using its own key agreement key $k_{ka}$, and the client's public key $k_{c+}$. 
It then decrypts the inputs and checks if the function hash $h(\mathcal{F})$ supplied by the client matches the hash of the loaded function.
If the hashes match, the function is invoked on the decrypted inputs.
If the function produces an output, this ECALL returns the size of the output, allowing the host application to allocate the correct size buffers outside the enclave.

\textbf{\texttt{finish:}} Finally, this ECALL copies the encrypted output and signed resource measurements out of the enclave.
These will be sent to the client and function provider respectively.

\section{SGX Resource Measurement}
\label{sec:resource_measurement}

\subsection{Compute Time}
\label{sec:timing}

By far the most challenging quantity to measure is the time spent executing the function's code within the enclave.
There are various pitfalls of performing this type of measurement in the presence of a potentially adversarial OS, as we assume in our threat model (Section~\ref{sec:threat_model}).
Specifically, we cannot rely on OCALLs and must account for the possibility that the OS could cause an asynchronous enclave exit (AEX) at any time.

\begin{figure}
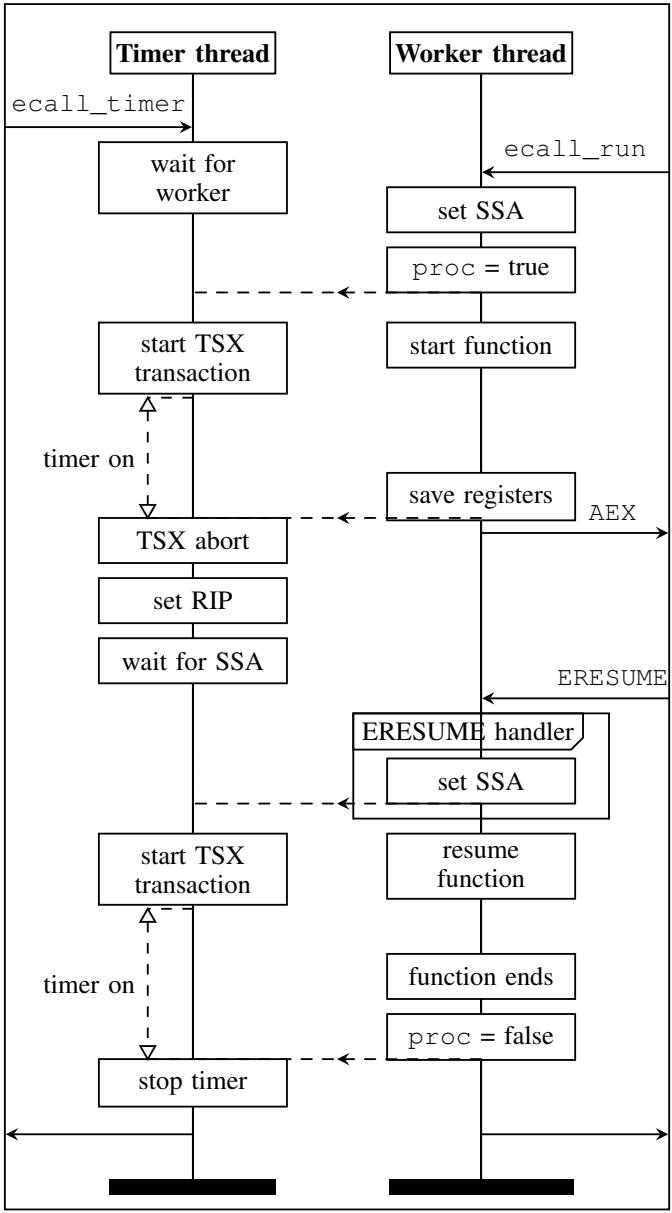


\setlength{\firstlevelheight}{3mm}

\setmsckeyword{}

\begin{msc}{}

\setlength{\instdist}{15mm}
\setlength{\envinstdist}{25mm}
\setlength{\topheaddist}{1mm}
\setlength{\actionwidth}{25mm}
\setlength{\actionheight}{6mm}
\setlength{\levelheight}{2mm}
\setlength{\bottomfootdist}{2mm}
\setlength{\inlineoverlap}{17mm}

    \declinst{t}{}{\textbf{Timer thread}}
    \declinst{w}{}{\textbf{Worker thread}}

    \nextlevel
    \nextlevel       
    \mess{\texttt{ecall\_timer}}{envleft}{t}
    \nextlevel
    \action{wait for worker}{t}
    \nextlevel
    \nextlevel
    \mess{\texttt{ecall\_run}}{envright}{w}
    \nextlevel
    \action{set SSA}{w}
    \nextlevel
    \nextlevel
    \nextlevel
    \nextlevel  
    \action{\texttt{proc} = true}{w}
    \nextlevel
    \nextlevel
    \nextlevel
    \order{w}{t}
    \nextlevel
    \nextlevel
    \action{start TSX transaction}{t}
    \action{start function}{w}
    \nextlevel
    \nextlevel
    \nextlevel
    \nextlevel
    \nextlevel
    \measure{timer on}{t}{t}[8]
    \nextlevel
    \nextlevel
    \nextlevel
    \nextlevel
    \nextlevel
    \action{save registers}{w}
    \nextlevel
    \nextlevel
    \nextlevel
    \order{w}{t}
    \action{TSX abort}{t}
    \nextlevel
    \mess{\texttt{AEX}}{w}[0.7]{envright}
    \nextlevel
    \nextlevel
    \nextlevel
    \action{set RIP}{t}
    \nextlevel
    \nextlevel
    \nextlevel
    \nextlevel
    \action{wait for SSA}{t}
    \nextlevel
    \nextlevel
    \nextlevel
    \nextlevel
    \mess{\texttt{ERESUME}}{envright}[0.3]{w}
    \nextlevel
    \inlinestart{ERES}{ERESUME handler}{w}{w} 
    \nextlevel
    \nextlevel
    \nextlevel    
    \action{set SSA}{w}
    \nextlevel
    \nextlevel
    \nextlevel
    \order{w}{t}
    \nextlevel
    \inlineend{ERES}
    \nextlevel
    \action{start TSX transaction}{t}
    \action{resume function}{w}
    \nextlevel
    \nextlevel
    \nextlevel
    \nextlevel
    \nextlevel
    \measure{timer on}{t}{t}[10]
    \nextlevel
    \nextlevel
    \nextlevel
    \action{function ends}{w}
    \nextlevel
    \nextlevel
    \nextlevel
    \nextlevel    
    \action{\texttt{proc} = false}{w}
    \nextlevel
    \nextlevel
    \nextlevel
    \order{w}{t}
    \action{stop timer}{t}
    \nextlevel
    \nextlevel
    \nextlevel
    \nextlevel
    \nextlevel
    \mess{}{t}{envleft}
    \mess{}{w}{envright}
    \nextlevel
\end{msc}
\caption{Overview of interactions between timer and worker threads.}
\label{fig:timing}
\end{figure}

We take a similar approach to the \emph{reference clock} proposed by Chen et al.~\cite{chen_detecting_2017}, but adapt this to measure resource usage.
Specifically, since all designs will inherently have some inaccuracy, we maintain the invariant that our timing mechanism will always provide a strict \emph{lower bound} of the time taken to execute the function, thus making it suitable for use in a billing policy.

The central idea is to use Intel's hyperthreading technology to enable \emph{two concurrently executing threads} per worker enclave: a \emph{worker} thread that executes the function, and a \emph{timer} thread that measures the time the worker thread spends inside the enclave.
Our timer thread measures time as an integer number of \emph{time periods}, each of duration \tperiod CPU cycles.
This is analogous to a clock where \tperiod is the duration between each clock tick.
The \tperiod CPU cycles are in turn measured using a calibrated \texttt{for} loop.
As we discuss in Section~\ref{sec:tperiod}, the value of \tperiod can be configured by the service provider, but will always be reported in the signed resource measurements. 

The main challenge is that the timer thread must be able to check whether the worker thread is currently in the enclave, especially given that the OS can interrupt the worker thread at any time.
This would be trivial to check if the timer thread could read the Thread Control Structure (TCS) of the worker thread, but in the current version of SGX, any explicit memory access to a TCS results in a Page-Fault exception~\cite{intel_sdm}.
To achieve this functionality with current hardware, we use Intel's Transaction Synchronization Extensions (TSX).
An overview of our approach is shown in Figure~\ref{fig:timing} and described in detail below.

\subsubsection{Start and end of function}

When starting a new invocation of a function, the host application spawns two threads.
With the first thread, it calls the \texttt{ecall\_timer} ECALL, thus making this the timing thread.
Once inside the enclave, the timing thread waits on a condition.
The host application then uses the second thread to call the \texttt{ecall\_run} ECALL, as described in Section~\ref{sec:worker_enclave}, making this the worker thread.
It is always in the service provider's interest to ensure that the timer thread has been started before starting the worker thread.

Once inside the enclave, the worker thread places a special \emph{marker value} in its current save state area (SSA).
This is similar to the approach used by both Cloak~\cite{gruss_strong_2017} and Varys~\cite{oleksenko_varys:_2018} to ensure two threads remain in the same enclave.
The worker thread then resets the \tcur variable to zero, sets the enclave-wide \texttt{proc} flag to indicate that processing has started, and signals to the timer thread via the condition.

The timer thread acquires an enclave-wide mutex to ensure that only one timer thread can be active at any given time, thus preventing a malicious service provider from double-counting the time by running multiple timer threads.
The timer thread checks that the \texttt{proc} flag is set, and if so enters a TSX transaction (which we refer to as the \emph{timer transaction}).
Within the transaction, the timer thread checks that the worker thread's SSA contains the marker value, which causes TSX to put the worker thread's SSA into the read set of the transaction.
The timer thread then executes a \texttt{for} loop to count for a number of CPU cycles, as defined based on \tperiod.
Once the loop completes, the timer thread increments its internal counter and exits the TSX transaction.
Immediately after the transaction, the timer thread copies its internal counter to \tcur.
Finally, the timer thread checks if the \texttt{proc} flag is still set, and if so begins a new timer transaction and repeats the above process.
On completion of the function, the worker thread clears the \texttt{proc} flag, reads the final time value $\tcur = \tmax$, and includes this in the authenticated resource measurement report.

In the above design, \tcur is essentially a mirror of the internal counter.  
The reason for using these two variables is to avoid making \tcur part of the transaction's write set, as would be the case if \tcur were incremented within the transaction.
If \tcur were part of the write set, operations that read \tcur during a transaction from another thread (e.g.\ our memory measurement mechanism) may under certain conditions cause the timer transaction to abort.

One potential problem faced by Chen et al.~\cite{chen_detecting_2017} in the design of their reference clock is that if the adversary can interrupt the clock thread between the end of the transaction and the increment of the clock variable (e.g.\ using the SGX-Step framework~\cite{van_bulck_sgx-step:_2017}), he can cause the clock to lose time.
However, this is not a problem in our scenario, because it is not in the service provider's interest to under-report the function's compute time.
In other words, interrupting our timer thread at this (or any other point) is always detrimental to the adversary.

\subsubsection{Worker thread interrupt}
\label{sec:worker_int}

In some cases, the OS may legitimately need to interrupt and resume the worker thread.
We therefore require a way for the timer thread to detect when the worker thread has been interrupted and pause the timer.
When the worker thread is interrupted (i.e.\ an AEX event), the CPU will save the registers to the worker thread's SSA.
This will cause the TSX transaction of the timer thread to abort, since the worker thread's SSA is in the timer transaction's read set.
If the transaction is aborted, neither the internal counter nor \tcur will be incremented for this partially completed tick.
When the timer thread attempts to restart the transaction, it will detect that the worker thread has been interrupted because the marker in the worker thread's SSA was overwritten by the CPU during the AEX.

Having detected the interrupt, we also need a way for the timer thread to detect when the worker thread has been resumed.
Typically, the ERESUME instruction transparently restores the CPU registers from the SSA and continues execution from the next instruction, which would not allow us to signal to the timer thread or recreate the marker in the SSA.
We overcome this challenge by creating a custom ERESUME handler inside the enclave.
Specifically, when the timer thread detects that the worker thread has been interrupted, it modifies the worker thread's SSA and swaps the instruction pointer (\texttt{RIP}) with the address of our custom handler.
The original instruction pointer value is saved separately so that it can be accessed by our handler.
Thus when the worker thread is resumed via ERESUME, it executes our custom handler instead of resuming the function.

Listing~\ref{lst:eresume_handler} shows the main functionality of our custom ERESUME handler.
We need the custom handler to recreate the marker in the worker thread's SSA and then continue executing the function from the point at which it was interrupted. 
However, since the ERESUME instruction has already restored all the registers from the SSA, our custom handler cannot clobber any registers.
We therefore implement the handler directly in assembly to control its precise behavior.
As shown in Listing~\ref{lst:eresume_handler}, we first store the \texttt{rax} and \texttt{rbx} registers on the stack.
We then dereference a pointer to the worker thread's SSA in order to write our marker value (\texttt{\$12345}).
Finally, we pop \texttt{rbx} and \texttt{rax} and jump to the original \texttt{RIP}.

\begin{lstlisting}[frame=single, language=C, caption={Custom ERESUME handler}, captionpos=b, float, frame=tblr, label=lst:eresume_handler]
.text
.globl custom_eresume_handler
.type custom_eresume_handler,@function
custom_eresume_handler:
  push   push   lea g_worker_ssa_gpr(  mov (  movl $12345,(  pop   pop   jmp *g_original_ssa_rip(
\end{lstlisting}

\subsubsection{Timer thread interrupt}
\label{sec:timer_int}

As explained above, interrupting the timer thread is always detrimental to the service provider because it reduces the measured compute time for the function.
However, it may still be necessary for an honest service provider to interrupt the timer thread.
If the timer thread was inside a transaction when it was interrupted, the transaction will be aborted.
When the timer thread is resumed, it will simply continue with the timing loop as described above (i.e.\ checking the \texttt{proc} flag and the marker value in the worker thread's SSA).
No custom resume handler is required for the timer thread.

In rare cases, the OS may also need to interrupt the worker thread while the timer thread is interrupted.
If the timer thread is resumed first, it will detect that the worker thread has been interrupted, based on the absence of the marker in the worker thread's SSA, and will set up the custom ERESUME handler, as described in Section~\ref{sec:worker_int}.
If the worker thread is resumed first, it cannot notify the (interrupted) timer thread, and so any compute time will not be measured until the timer thread is resumed and the worker thread is interrupted and resumed again.
Neither of these scenarios allow the service provider to inflate the compute time measurement.
As before, it is always in the service provider's interest to ensure that the timer thread is running before starting or resuming the worker thread.

\subsubsection{Worker thread OCALL}

If the worker thread is required to perform an OCALL, we need to pause the timer thread and resume it when the OCALL returns.
We instrument the OCALL code to clear the enclave-wide \texttt{proc} flag.
This does not interrupt the timer thread (i.e.\ it will complete the current tick), but prevents it from counting further.
Although it would be possible to interrupt the timer thread, this could be abused by a malicious function provider triggering frequent OCALLs in an attempt to reduce the measured compute time.
When the OCALL returns, it sets the \texttt{proc} flag and signals to the timer thread to resume.

\subsubsection[Choosing T]{Choosing \tperiod}
\label{sec:tperiod}

We allow the service provider to choose \tperiod independently for each worker enclave.
This parameter cannot be changed once the enclave has been initialized, and it is included in the authenticated resource measurement report provided at the end of the function.
Making this parameter configurable allows the service provider to determine the optimum value for their environment, based on their expected interrupt frequency.
A shorter value of \tperiod would reduce the amount of time under-reported if the worker or timer thread is interrupted during the \texttt{for} loop.
However, the timer thread must perform various checks between transactions, which are not included in the measured time.
A shorter value of \tperiod would result in more frequent transactions and thus more cycles spent on these checks, also leading to under-reporting.

\subsection{Memory Usage}
\label{sec:memory_usage}

The most fine-grained measurement of a function's memory usage is the time-integral \mint of the function's memory allocation over the duration of the function:
\begin{equation}
\mint = \int_{0}^{\tmax}\mem\ \mathrm{d}t
\end{equation}

where \tmax is the total compute time of the function, and \mem is the instantaneous memory allocated by the function at time \tcur.
We also measure the maximum instantaneous memory usage at any point during the function's execution (\mmax), and the average memory usage over the duration of the function (\mavg):
\begin{align}
\mmax & = \max_{0 \leq t \leq \tmax} \mem \\[1.5ex]
\mavg & = \dfrac{\mint}{\tmax}
\end{align}

The service provider can use either or both measurements in their billing policy.
To measure memory usage, we instrument the \texttt{malloc}, \texttt{realloc}, and \texttt{free} functions used by the sandbox.
This ensures that any code running within the sandbox can only use these instrumented functions.
As memory is allocated, reallocated, or freed, we update \mint and \mmax.
We use the internal variable \mem to represent the current amount of memory allocated by the function at time \tcur, and the variable \tmem to represent the time at which \mem was last changed.

On any \texttt{malloc}, \texttt{realloc}, or \texttt{free} event, we perform Algorithm~\ref{alg:memory}, with the net amount of memory allocated or freed as the input \deltam.
Specifically, we first obtain the current value of \tcur and subtract \tmem from this to obtain the number of time periods since the previous change \deltat.
We then multiply \mem by \deltat and add the result to \mint.
We then increase or decrease \mem by the amount of memory allocated or freed, and set \tmem to the current value of \tcur.
Finally, we check if the new \mem value exceeds the previous maximum value \mmax and, if so, we update \mmax.

\begin{algorithm}[H]
\caption{Memory measurement update}
\label{alg:memory}
\begin{algorithmic}[1]
\renewcommand{\algorithmicrequire}{\textbf{Input:}}
\renewcommand{\algorithmicensure}{\textbf{Output:}}
\REQUIRE $\deltam$

\STATE $\deltat \gets \tcur - \tmem$
\STATE $\mint \gets \mint + (\deltat \times \mem)$
\STATE $\mem \gets \mem + \deltam$
\STATE $\tmem \gets \tcur$

\IF {$(\mem > \mmax)$}
  \STATE $\mmax \gets \mem$
\ENDIF

\end{algorithmic}
\end{algorithm}

Since the function may terminate without explicitly freeing all its allocated memory, we perform a final integration step at the end of the function: we calculate the \deltat since the last \tmem, multiply this by the final value of \mem, and add the result to \mint.

Since our timing mechanism counts time in integer multiple of \tperiod (i.e.\ ticks), we cannot measure time intervals smaller than \tperiod CPU cycles.
In Algorithm~\ref{alg:memory}, this means that any memory changes \deltam occurring \emph{between} ticks are all treated as having occurred at the last tick.
Depending on the behavior of the function, this could result in slightly over- or under-reporting \mint.
For example, memory that is allocated just before a tick will be treated as having been allocated nearly one tick earlier.
Conversely, memory that is allocated and freed between two ticks will not be included in \mint.
This is an inherent limitation of using a time source with values of $\tperiod > 1$.
However, note that the value of \mmax will always be precisely calculated, so if a function allocates and then frees a very large memory area between ticks, this will be visible in \mmax.
We quantitatively evaluate this in Section~\ref{sec:evaluation_accuracy}, and show that memory measurement accuracy can be maximized by setting a small value of \tperiod.

\subsection{Network Usage}
\label{sec:network_usage}

Networking calls from inside the enclave are passed to the untrusted environment via an OCALL.
After the OCALL resumes, the total number of bytes sent and received is added to the network usage metric before the output of the networking call is returned to the function.

\section{Implementation in OpenWhisk}
\label{sec:openwhisk}

We integrated our \name architecture and resource measurement mechanisms in the Apache OpenWhisk FaaS framework~\cite{openwhisk} in order to 1) demonstrate that our architecture is compatible with an existing framework, and 2) evaluate the overall performance impact of our changes.
OpenWhisk supports functions written in various languages, including JavaScript/NodeJS, Swift, Python, and Java.
It also supports custom logic in Docker containers.

To integrate \name with OpenWhisk, we created a new Docker image containing our worker enclave.
Inside our worker enclave we implemented the resource measurement mechanisms described in the previous section.
Similarly to Milutinovic et al.~\cite{milutinovic_proof_2016}, we use the Duktape JavaScript interpreter~\cite{duktape} because of its portable design and low memory footprint.
However, we could have used any suitable interpreter or sandbox, including the \emph{MuJS} JavaScript interpreter used by Goltzsche et al.~\cite{goltzsche_trustjs:_2017} in the TrustJS system, or Ryoan~\cite{hunt_ryoan:_2016}, a request-oriented sandbox for SGX.

When a client submits a request, OpenWhisk initially stores this in a database of pending requests until it can be dispatched to a suitable worker.
In \name, this input is already encrypted, as explained in Section~\ref{sec:architecture}, but this does not require modification of OpenWhisk.
To handle this request, OpenWhisk creates a new Docker container, which initializes the worker enclave.
As explained in Section~\ref{sec:architecture}, if a sealed key set is not already available, the worker enclave contacts the provisioning enclave and requests the necessary keys.
Once a key set has been provisioned to a particular platform, we store the sealed key set locally and make this available to all Docker containers on the platform.
The worker enclave then processes the request as usual, and produces the output (if any) encrypted for the client.
OpenWhisk again stores this output in a database until it is sent to the client.
The signed resource measurements produced by our worker enclaves can also be stored in the OpenWhisk database and sent to the function provider for billing purposes.

\section{Evaluation}
\label{sec:evaluation}

\subsection{Security Analysis}
\label{sec:evaluation_securityy}

\textbf{Security:}
To meet Requirement~\textbf{R1}, a malicious service provider must be prevented from modifying the inputs or outputs of a function invocation, and the client must receive assurance that received output $\mathcal{O}$ is the result of a correct execution of the intended function $\mathcal{F}$ on the supplied inputs $\mathcal{I}$.
As explained in Section~\ref{sec:transitive_attestation}, our transitive attestation protocol provides the equivalent level of assurance as if the client had directly attested the specific worker enclave.
After attesting the key distribution enclave (KDE), the client can therefore trust that only a worker enclave with the correct MRENCLAVE value will have access to the key set published by the KDE.
The use of authenticated encryption protects both the confidentiality and integrity of the client's inputs in transit.
Even if the service provider provisions an incorrect function, the worker enclave will detect the mismatch based on the hash of the intended function included in the client's input.
If the service provider replays a previous output to the client, the client will detect this because the nonce in the output will not match the nonce the client included in the encrypted input.
Even if the client accidentally re-uses nonces, any manipulation of $\mathcal{I}$, $\mathcal{F}$, or $\mathcal{O}$ would be visible in the signed receipt. 
Similar arguments apply for the function provider establishing the trustworthiness of a worker enclave and checking the veracity of a resource measurement report produced by the enclave (Requirement~\textbf{R4}).

\textbf{Input \& output privacy:}
Given the known side-channel attacks against SGX, we cannot guarantee the confidentiality of inputs and outputs for arbitrary functions.
Functions that exhibit secret-dependent control flow are likely to be vulnerable to attacks that infer this control flow from outside the enclave (e.g.~\cite{lee_inferring_2017}) whilst those exhibiting secret-dependent memory access could be vulnerable to cache and page-based side-channel attacks (e.g.~\cite{xu_controlled-channel_2015, bulck_telling_2017, brasser_software_2017}).
Although it would be possible to include various defenses against these some of these side-channel attacks (e.g.~\cite{chen_racing_2018, chen_detecting_2017, gruss_strong_2017, seo_sgx-shield:_2017}), this is an orthogonal problem to ours.
Notably, three state of the art defenses against cache side-channel attacks, Cloak~\cite{gruss_strong_2017}, HyperRace~\cite{chen_racing_2018}, and Varys~\cite{oleksenko_varys:_2018}, all require that the attacker is prevented from controlling the sibling hyperthread.
Our timer thread inherently fulfils this requirement.
Furthermore, since our worker enclave itself does not exhibit secret-dependent control flow or memory access patterns, we do not introduce any vectors for side-channel attacks that were not already present in the provisioned function.
Therefore we can only achieve requirement \textbf{R2} for functions exhibiting secret-independent control flow and memory access patterns.

\textbf{Resource measurement integrity:}
As explained in Section~\ref{sec:architecture}, the sandbox in the worker enclave protects the integrity of our resource measurement mechanisms against the function.
Therefore, even a malicious function provider cannot interfere with the timing thread, evade the instrumented memory management functions, or modify the internal state of any of these mechanisms.
Conversely, the enclave protects the resource measurement mechanisms against a malicious service provider.
Although the service provider can always interrupt either the timer or worker thread, this is never in the service provider's interest, as explained in Section~\ref{sec:resource_measurement}.
Therefore, \name{} fulfils Requirement~\textbf{R4}.

\textbf{TCB size:}
We measured the size of the Trusted Computing Base (TCB) of the worker enclave and key distribution enclave (KDE) using David A. Wheeler's \texttt{sloccount} tool.\footnote{\url{https://www.dwheeler.com/sloccount/}}
The Duktape JavaScript interpreter itself is approximately 65,000 lines of C code.
However, we assume this code is trustworthy, and could instead use other types of sandboxes (e.g.\ Ryoan~\cite{hunt_ryoan:_2016}) in order to minimize the TCB.
Excluding the SGX trusted libraries and the JavaScript interpreter, our resource measurement mechanisms add approximately 731 lines of C++ code to the worker enclave.
Our KDE consists of only 195 lines of C++ code.
Thus the critical \name{} code is amenable to security audits or possibly even formal verification.

\textbf{Damage containment:}
By default, all worker enclaves share the same MRENCLAVE value, so if one instance of a worker enclave were compromised, it would be able to unseal keys used for different functions in other worker enclaves.
To mitigate this risk, \name{} can be configured to provide \emph{damage containment} by parametrizing the worker enclaves, either by function or by client.
Since this parametrization changes the worker enclave's MRENCLAVE value, the enclave is restricted to certain functions of clients.
Data sealed by a parametrized enclave is thus protected against enclaves with different parametrization or non-parametrized enclaves.
The service provider can thus use this mechanism to balance between flexibility (i.e., the ability to run any task on any worker enclave) and damage containment. 
The parametrization can also be applied selectively, e.g., creating a set of parametrized worker enclaves for each large customer, and a set of general-purpose worker enclaves for other customers.

\subsection{Measurement Accuracy}
\label{sec:evaluation_accuracy}

We evaluate the accuracy of our compute time, memory, and network usage measurements by running synthetic benchmark functions with known characteristics.
To evaluate compute time and memory, we use a synthetic \texttt{Fibonacci} function that takes a single integer $n$ as an input parameter.
This function calculates the first $n+1$ numbers in the Fibonacci sequence and stores them in a pre-allocated list, as shown in Listing~\ref{lst:fibonacci}.
Both the time and memory complexity of this function are therefore $O(n)$ in the input parameter.

\begin{lstlisting}[frame=single, language=JavaScript, caption={Fibonacci test function}, captionpos=b, float, frame=tblr, label=lst:fibonacci]
function input(params) {
    var n = params.iterations || 1;
    var values =  new Array(n);
    values[0] = 0;
    values[1] = 1;
    
    for(var i=2; i<=n; i++){
        values[i] = values[i-1] + values[i-2];
    }
    
    var udata = params.udata || "udata"
    var result = JSON.stringify({result: values[n]});
    return JSON.stringify({output:result, 
        measurements_udata: udata});
}
\end{lstlisting}

\begin{figure}
\includegraphics[trim={25mm 10mm 30mm 22mm},clip,width=\columnwidth]{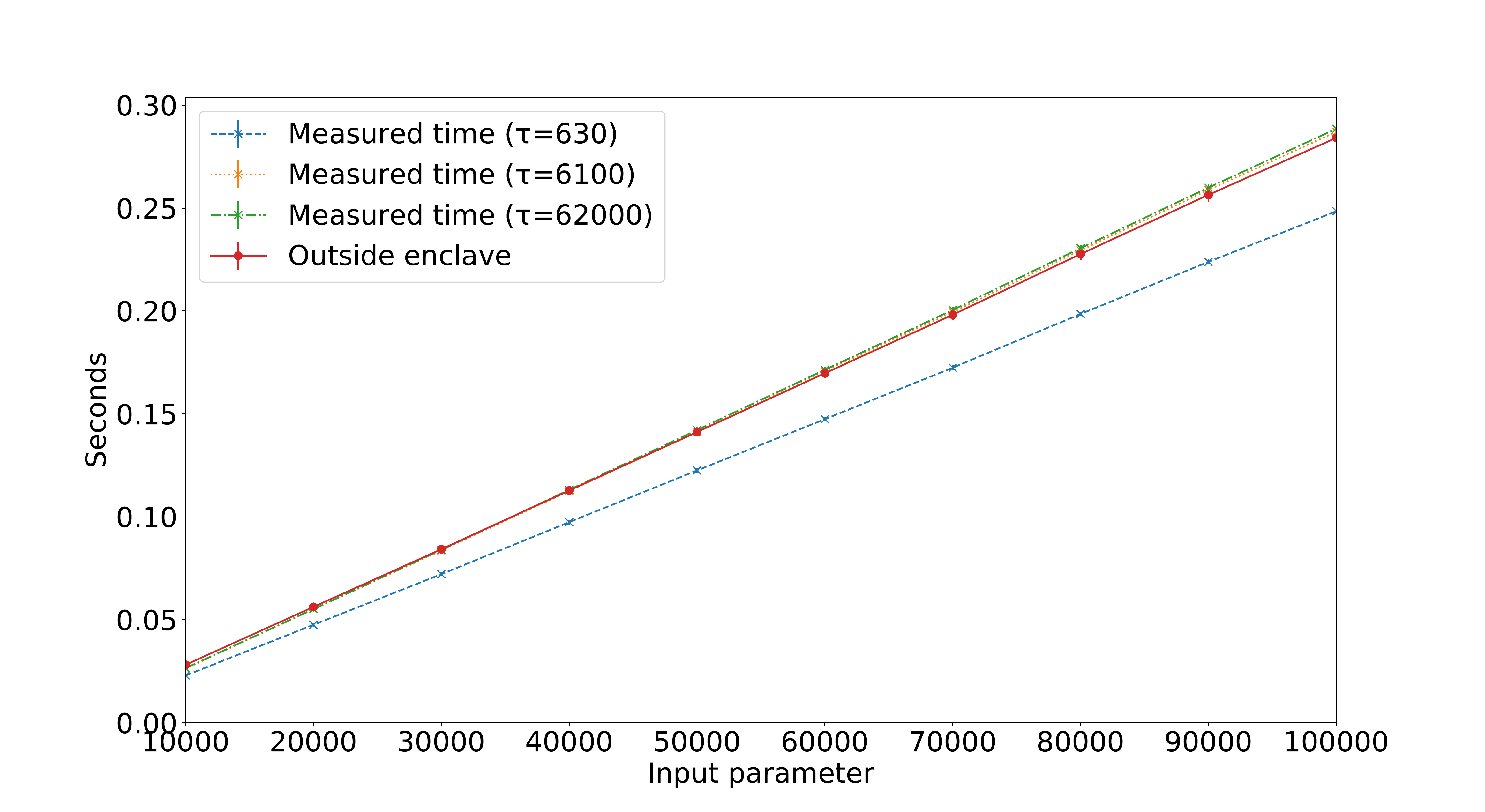}
\caption{Compute time (\tmax) measurements of the \texttt{fibonacci} function for different values of \tperiod (average of 10 runs).}
\label{fig:timing_accuracy}
\end{figure}

Figure~\ref{fig:timing_accuracy} shows the measured time (y-axis) for different values of the input parameter (x-axis), with different values of \tperiod.
Although we did not explicitly interrupt the worker thread, but the normal OS scheduling still caused some interrupts during these tests.
For comparison, we also measured the total runtime of the \texttt{ecall\_run} ECALL from outside the enclave, indicated as \emph{outside enclave} in Figure~\ref{fig:timing_accuracy}.
Although the time measured outside the enclave also includes the ECALL, the time taken for this is negligible in comparison to the runtime of the function.
Given that the time complexity of our \texttt{Fibonacci} function is $O(n)$ in the input parameter, it would be expected that the time increases linearly with the input parameter.
As shown in Figure~\ref{fig:timing_accuracy}, for all values of \tperiod our measured values exhibit linear behavior in the input parameter.
As expected, setting a very low value of $\tperiod = 630$ cycles results in under-reporting, as explained in Section~\ref{sec:tperiod}.

\begin{figure}
\includegraphics[trim={25mm 10mm 30mm 22mm},clip,width=\columnwidth]{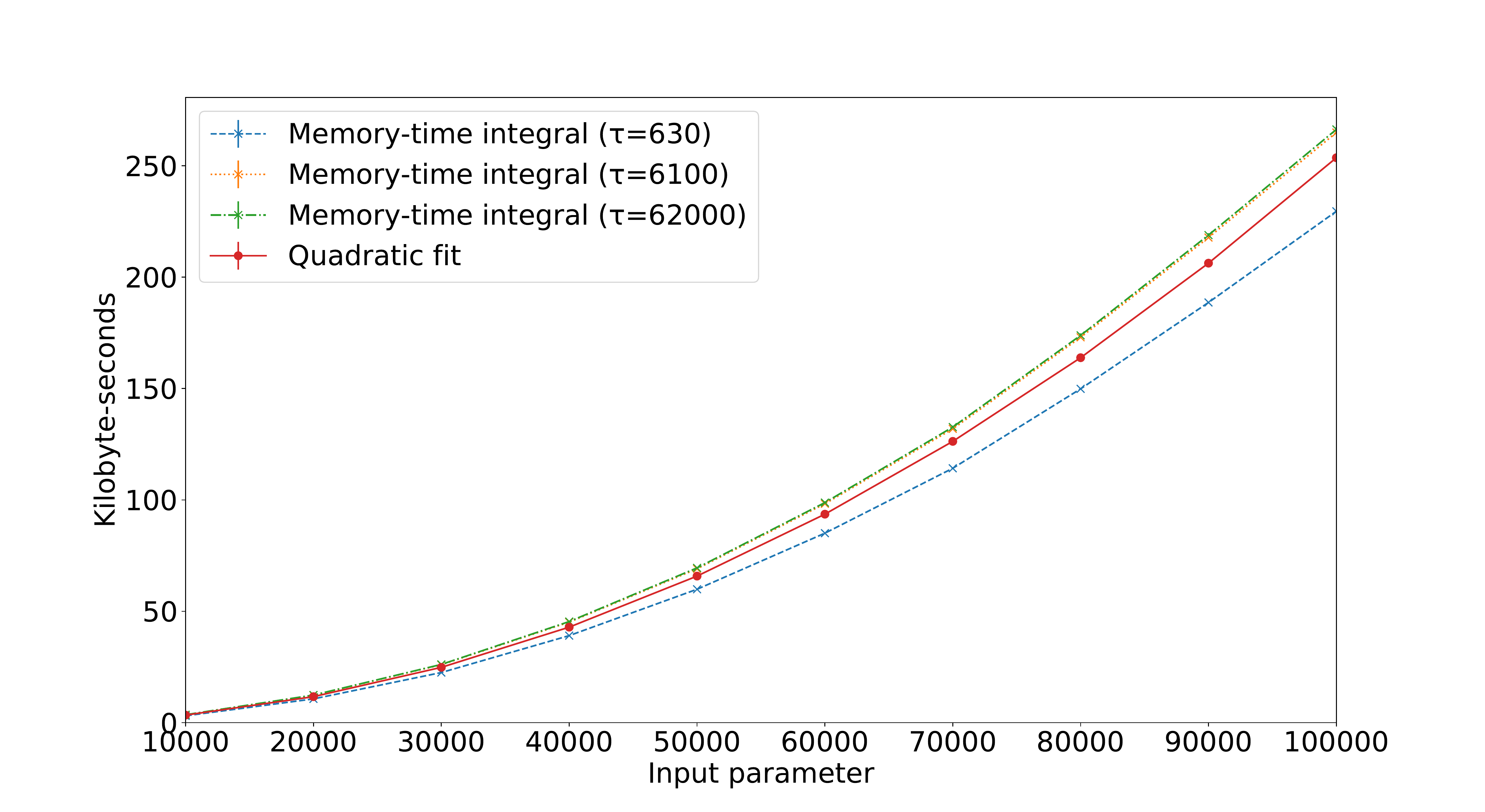}
\caption{Memory time-integral (\mint) measurements of measurements of the \texttt{fibonacci} function for different values of \tperiod (average of 10 runs).}
\label{fig:memory_accuracy}
\end{figure}

Figure~\ref{fig:memory_accuracy} shows the average values of \mint for different values of the input parameter.
As before, we did not explicitly interrupt the worker thread,
Given that both the time and memory complexity of our \texttt{Fibonacci} function are $O(n)$ in the input parameter, it would be expected that the time-integral of memory \mint increases quadratically with the input parameter.
As shown in Figure~\ref{fig:memory_accuracy}, our measured values all exhibit quadratic behavior in the input parameter (cf. the dashed quadratic line in the figure).
As with the timing measurement, very low values of \tperiod result in under-reporting. 
Although it is not straight-forward to perform the same type of measurement from outside the enclave, we confirmed the validity of our \mint measurements by dividing the measured \mint values by \tmax to obtain the time-averaged memory usage for each function, and checked that this increases linearly in the input parameter, as expected.

Finally, to test network usage, we created a synthetic \texttt{known\_network} function that sends and receives a specified number of bytes via the network interface.
In all cases, the enclave's resource measurement mechanism correctly measured the number of bytes sent and received.
Our resource measurement mechanisms therefore fulfil Requirement~\textbf{R3}.

\subsection{Performance}
\label{sec:evaluation_performance}

We quantified two different aspects of the additional latency incurred when using \name{}: 1) the impact of our \name{} resource measurement mechanisms, and 2) the overall impact of \name{} when integrated with OpenWhisk.
The former depends on the specific function, whilst the latter is a fixed latency overhead that applies to any function.
All benchmarks were built with the Intel SGX SDK version~2.2 and run on Ubuntu~18.04 on an Intel Core~i5 CPU with 8~GB of RAM.

\textbf{Resource Measurement Latency:}
To quantify the latency increase of our resource measurement mechanisms on the function, we used several benchmark functions from the Octane~\cite{octane} JavaScript benchmark suite.
This is the same benchmark suite used by the developers of the Duktape interpreter to compare performance between versions and other interpreters.
Although the Octane benchmarks include their own timing functionality, we do not use this because we want to measure the overall execution time.
We excluded some benchmarks that are not realistic use cases of FaaS (e.g.\ displaying graphics).
The functionality of each benchmark is described in the Octane benchmark reference~\cite{octane}.

We ran each benchmark for three different scenarios.
In the first scenario, the inputs and outputs are not encrypted.
This would be used during incremental deployment of \name{} (see Section~\ref{sec:deployability}), or for clients who do not require encryption.
In the second scenario, the inputs and outputs are encrypted, and in the third scenario, the client additionally requests a signed receipt of the function's execution, as explained in Section~\ref{sec:architecture}.
In all cases, \name{} provides a signed resource measurement report.
As a baseline, we compare these measurements against the same Duktape interpreter, running inside an SGX enclave, with no input/output encryption or resource measurement (i.e.\ the same environment as used by Milutinovic et al.~\cite{milutinovic_proof_2016}).
All measurements are the average of 10 runs.

\begin{table}
\renewcommand*{\arraystretch}{1.3}
\caption{Resource measurement overhead}
\label{tab:resource_measurement_performance}
\begin{tabularx}{\columnwidth}{|X|c|c|c|c|}
\hline
\textbf{Function} & \textbf{Baseline} & \multicolumn{3}{c|}{\textbf{\nameVarSize{}}} \\
\hline
                  & Time (seconds)    & No encryption      & Encryption   & Enc. \& receipt \\
\hline
Box2D             & 3.019             & 3.118              & 3.121        & 3.135 \\
DeltaBlue         & 1.446             & 1.524              & 1.529        & 1.537 \\
NavierStokes      & 4.155             & 4.418              & 4.447        & 4.473 \\
RayTrace          & 0.779             & 0.848              & 0.850        & 0.852 \\ Richards          & 1.719             & 1.767              & 1.767        & 1.799 \\ \hline
Mean              & 1.893             & 1.993              & 1.998        & 2.013 \\
\hline
Overhead          &   -               & 5.3\%              & 5.6\%        & 6.3\% \\
\hline
\end{tabularx}
\end{table}

As shown in Table~\ref{tab:resource_measurement_performance}, the average overhead of our \name{} resource measurement over the benchmark suite ranges from 5.3\% without encryption to 6.3\% with full encryption and a signed transaction receipt.
The resource measurement overhead includes any additional time required to initialize and synchronize the worker and timer threads and to generate the signed resource measurement report.

\textbf{Pre-function latency:}
To quantify the additional end-to-end latency of starting a function when \name{} is integrated with OpenWhisk, we measured the response time of an empty JavaScript function with and without \name{} in both cold-start and warm-start scenarios.
In the \emph{cold-start} case, there are no \name{} Docker containers or Worker Enclaves running, as would be the case for the first invocation of a function.
This measurement therefore includes the time to start the Docker container, initialize the Worker Enclave, unseal the key set, load the Duktape interpreter, provision the function to the interpreter, marshal the inputs into the Worker Enclave, perform the key agreement with the client, decrypt the inputs, and return the result to the client.

For frequently used functions, it is likely that the system would already have a Docker container and Worker Enclave loaded for the specific function, resulting in a significantly faster \emph{warm-start}.
In this case, the measurement includes only the time to marshal the inputs into the Worker Enclave, perform the key agreement with the client, decrypt the inputs, and return the result to the client.
Since we are using an empty function, these measurements are independent of the specific function invoked, and thus constitute a fixed latency overhead of using \name{} with OpenWhisk.
As a baseline, we compare these measurements against the same Duktape JavaScript interpreter, running inside a native OpenWhisk container.
All measurements are the average of 10 runs.
As shown in Table~\ref{tab:prefunction_latency}, \name{} adds less than 3\% additional latency before the function.

\begin{table}
\renewcommand*{\arraystretch}{1.3}
\caption{Pre-function latency for OpenWhisk + \nameVarSize{}}
\label{tab:prefunction_latency}
\begin{tabularx}{\columnwidth}{|X|r r|r r|c|}
\hline
                   & \textbf{Baseline} & (std dev) & \textbf{\nameVarSize{}} & (std dev)  & Overhead \\
\hline
Cold start         & 3,179~ms          & (40~ms)   & 3,246~ms                & (38~ms)    & 2.1\%    \\
Warm start         &   204~ms          & (106~ms)  &   210~ms                & (149~ms)   & 2.9\%    \\
\hline
\end{tabularx}
\end{table}

All benchmarks in this section assume an unloaded server, as may be the case for a small ephemeral service provider.
As future work, we plan to evaluate the performance of \name{} on a fully-loaded cluster, as might be found in a large cloud provider's data center.
We could use a similar approach to Vaucher et al.~\cite{vaucher_sgx-aware_2018}, who performed this type of evaluation for SGX-enabled containers in the Kubernetes orchestrator using the Google Borg traces.

\section{Discussion}
\label{sec:discussion}

\subsection{Trade-offs and Limitations}

\textbf{Additional thread:}
One potential limitation of our mechanism for measuring compute time is that, for each worker thread, we have to dedicate the sibling hyperthread to solely perform the timing measurement.
However, there are two reasons why our approach will \emph{not} significantly diminish performance in real-world deployments.
Firstly, since the sibling hyperthreads share the first level (L1) cache, it is known that some cloud providers disable hyperthreading to prevent sibling hyperthreads from causing costly cache evictions.
Secondly, controlling both sibling hyperthreads is already necessary for preventing various cache-line side-channel attacks, as is the case in Cloak~\cite{gruss_strong_2017}, HyperRace~\cite{chen_racing_2018}, and Varys~\cite{oleksenko_varys:_2018}.
Our timer thread can thus serve as a benign sibling hyperthread.

\textbf{Timing granularity:}
As explained in Section~\ref{sec:tperiod} and Section~\ref{sec:memory_usage}, the granularity of our timing mechanism is an inherent trade-off.
Reducing the duration of the period \tperiod would improve the accuracy of the compute time measurement \tmax, and the time-integral of memory usage \mint.
However, shorter periods require more frequent TSX transactions, thus increasing the percentage of time the timer thread spends creating these transactions (i.e.\ thus under-reporting \tcur and \mint).
We therefore allow each service provider to determine the optimum value of \tperiod for their specific circumstances and billing policy, whilst still ensuring the trustworthiness of the resource measurement by including \tperiod in the signed resource measurement.

\textbf{Architecture-specific calibration:}
In our timing thread, we have calibrated the iteration count of the \texttt{for} loop to take the specified number of CPU cycles (\tperiod) on \emph{current} SGX-capable CPUs.
Although this suffices for all current SGX parts, we may need to revisit this calibration for future SGX-enabled CPUs.
This would also require a trustworthy mechanism through which the service provider could report what type of CPU and timing calibration they are using.

\subsection{Suggested SGX Enhancements}

Given the challenges we faced in designing and implementing \name{}, we propose a \emph{wish list} of improvements to the current SGX design.
We only include suggestions that do not overtly require significant changes to the current SGX design.

\textbf{Secure tick count:}
As explained in \mbox{Section~38.6.1} of the Intel Software Developer's Manual~\cite{intel_sdm}, reading the CPU tick counter using the \texttt{RDTSC} instruction is only supported in SGX2.
However, even if accessible, this cannot be used for resource measurements because its value can be manipulated by privileged system software outside the enclave (e.g.\ the hypervisor can virtualize this tick counter and provide different values to different VMs).
If SGX enclaves had access to a \emph{secure} tick counter that could not be manipulated by untrusted software, \name{} could use this instead of our calibrated loop.

\textbf{ERESUME handler:}
Even with a secure tick counter, the timing mechanism would still need a way to detect asynchronous enclave exists and transparent ERESUME events.
If the enclave could specify a custom ERESUME handler, similar to our design in section~\ref{sec:worker_int}, this could be used to account for these events.
To accurately measure time spent inside the enclave, this handler would still need to know when the enclave was interrupted.
This could be achieved by having the CPU store the current value of the secure tick counter in the thread's SSA when the enclave is interrupted.
Upon ERESUME, the custom handler could simply calculate the time for which the enclave was interrupted.
This custom ERESUME handler could also be used to enhance the security of enclaves in other ways.
For example, it could be used to directly detect unusually frequent enclave interrupts, similarly to the D\'ej\`a~vu system by Chen et al.~\cite{chen_detecting_2017}.

Concurrently with our work, Oleksenko et al.~\cite{oleksenko_varys:_2018} proposed similar SGX hardware enhancements to improve the efficiency of \emph{Varys}, their system for defending against side-channel attacks on SGX enclaves.
This indicates that the above enhancements could benefit multiple different types of systems.

\subsection{Deployability Considerations}
\label{sec:deployability}

\textbf{Deployment without client changes:}
To improve deployability, \name{} can be deployed in an incremental manner such that it does not require changing the service provider and clients simultaneously.
By making message encryption optional in the worker enclaves, \name{} can be deployed completely transparently to clients.
However, the function provider can still receive an authenticated report of the functions resource consumption.
Without client encryption, the function provider may require an alternative solution to check that the function is only called by authorized clients (e.g.\ a single-use authorization token bound to the client's inputs).
Once \name{} is in place, clients can incrementally transition to using encryption as needed.

\textbf{Implementations with other TEEs:}
Although our current implementation of \name{} uses Intel SGX and TSX, the design can in principle also be implemented using other TEE technologies that provide isolated execution and remote attestation.
For example, in ARM TrustZone, the \emph{secure world} is more privileged than the \emph{normal world}, and thus the former cannot be interrupted by the latter.
Running \name{} in the secure world would still require a suitable sandbox to constrain the function, but could potentially use a simpler timing approach if the secure world has access to a secure tick counter.
Although TrustZone itself does not provide remote attestation, this functionality is typically provided by the vendor of the secure world's trusted OS. 
Brenner et al.~\cite{brenner_trapps:_2017} have proposed \emph{TrApps}, an architecture for securing general-purpose cloud workloads using ARM TrustZone, which could make use of our resource measurement mechanisms.

\textbf{Integration with Smart Contracts:}
Various approaches have been proposed to improve security, privacy, performance, and efficiency of smart contracts by performing certain computations inside TEEs~\cite{bowman_private_2018, microsoft_coco, cheng_ekiden:_2018, kaptchuk_giving_2017}.
However, none of these describe how the service providers operating the TEEs are compensated for the use of their (often scarce) computational resources.
More broadly, the idea of using smart contracts to pay for outsourced computation has only recently begun to be explored.
For example, the Golem network~\cite{golem_network} uses Ethereum-based transactions to settle payments between users and providers of outsourced computation. 
However, it is unclear how \emph{accountability} is achieved in this setting.

Since our \name{} resource measurements can be automatically verified, these can be used directly in smart contracts to enable decentralized payment for outsourced computation.
For example, the billing policy could be instantiated as a smart contract that transfers funds to the service provider at a pre-agreed rate after presentation of a correctly signed and attested function receipt.
Using this type of smart contract without verifiable resource consumption measurements could lead to over-charging by service providers, or disputes that must be manually resolved.

Automated decentralized payment is particularly important for allowing small ephemeral service providers (e.g.\ individuals) to enter the market.
Since these small entities cannot in general rely on the reputation-based trust enjoyed by established cloud providers, \name{} enables them to prove that they will perform the computation securely and measure resource usage correctly.

\textbf{Other use cases:}
In addition to automated decentralized billing, verifiable resource usage measurements can also be used for other purposes.
For example, Zhang et al.~\cite{zhang_rem:_2017} and the \emph{Anker network}~\cite{ankr_network} have both proposed using TEEs to perform \emph{useful computations} as a mechanism for leader elections (e.g.\ to decide which node will mine the next block in a blockchain).
The more computations a miner performs, the greater their chance of mining the next block.
\name{} could also be used for this purpose.

\section{Related Work}

\textbf{Resource measurement:}
Tople et al.~\cite{Tople_vericount:_2018} proposed Vericount, a system for measuring resource usage of SGX enclaves.
Similarly to \name, Vericount places the resource measurement mechanism within the same enclave as the code to be measured, and measures compute time, memory, network bandwidth, and I/O resources used by the enclave.
It protects the resource measurement code from other code within the enclave using software fault isolation (SFI) techniques.
However, Vericount cannot be directly used in the FaaS context for two reasons:
firstly, given our adversary model, Vericount's mechanism for measuring compute time could be arbitrarily inflated by an adversarial service provider, and secondly its mechanism for measuring memory usage is too coarse-grained for the FaaS context.

To measure compute time, Vericount instruments every ECALL to read and store the starting time using the SGX trusted time function (\texttt{sgx\_get\_trusted\_time}) provided by the SGX SDK.
Before the ECALL returns, the same function is again called to obtain the end time.
Additionally, the start and end times of each OCALL are also recorded using the same timing API.
The total time spent inside the enclave can thus be calculated from the start and end times of the ECALL, minus the time spent on OCALLs.
In our adversary model, the service provider could interrupt the enclave \emph{without} causing an OCALL.
This type of interrupt triggers an Asynchronous Enclave Exit (AEX), from which the enclave can later be resumed using the ERESUME instruction.
However, since AEX events are not accounted for in Vericount, the time for which the enclave is interrupted will be included in the total CPU time.
Furthermore, the SGX trusted time function itself causes an OCALL, since the time value is provided by an architectural enclave.
Once control has been transferred to the malicious OS through this OCALL, the OS can delay the request arbitrarily.
For example, the OS can delay the final call to the SGX trusted time function to arbitrarily inflate the end time value.

To measure memory usage, Vericount only records the maximum \emph{allowed} memory of the enclave, but does not measure the enclave's \emph{actual} memory allocation, either as a maximum, average, or time-varying quantity. 
This is likely too coarse-grained for the FaaS context, because the memory requirements of a single function may vary significantly depending on e.g.\ the size of the input.
Since the enclave's maximum permissible memory is statically defined in the compiled enclave, a function provider would have to either pay for the largest possible memory allocation for every function invocation, or provide multiple enclaves with different maximum memory sizes, which would result in different MRENCLAVE values in the remote attestation.

\textbf{Measuring time:}
One solution to the timing problem is the \emph{reference clock} designed by Chen et al.~\cite{chen_detecting_2017} as part of the D\'ej\`a~vu system.
They were the first to suggest using TSX to measure time within an enclave, and they use this to measure the time between selected basic blocks of the enclave's code.
An unusually long execution time between basic blocks indicates that one or more AEXs have occurred, and this can be used to detect side-channel attacks.
In comparison to our resource measurement mechanism, the main difference is that D\'ej\`a~vu aims to implement an accurate \emph{clock} whereas we aim to implement an accurate \emph{lower bound timer} that can be used as the basis for billing policies.
This gives rise to two fundamental design differences:
Firstly, for the reference clock, it is important to detect interruption of the reference clock thread, which they achieve by randomizing the number of CPU cycles each transaction takes.
In contrast, we use a calibrated fixed \tperiod, so that this can be reported to the function provider.
As we explain in Section~\ref{sec:resource_measurement}, it is not in the service provider's interest to interrupt our timing thread.
Secondly, whereas D\'ej\`a~vu instruments selected basic blocks to read the reference clock when they are executed, we use TSX to detect when the worker thread is interrupted and pause our time measurement.

Liang et al.~\cite{liang_bring_2018} proposed \emph{TrustedClock} for SGX, which uses System Management Mode (SMM) to generate and sign a high-precision timestamp, so that it can be verified by the enclave.
However, since this involves an OCALL to pass the request from the enclave to SMM, this timing mechanism is not suitable for resource measurement because the untrusted OS could delay this OCALL, and thus arbitrarily inflate the measured time.

\textbf{Counting instructions:}
Instead of measuring time in CPU cycles, Zhang et al.~\cite{zhang_rem:_2017} propose to count the number of instructions executed within the enclave.
A core aspect of their Resource Efficient Mining (REM) framework is \emph{Proof of Useful Work} (\emph{PoUW}), in which a TEE is used to perform arbitrary useful computations.
The more computations a miner performs, the greater their chance of mining the next block.
To count the number of executed instructions, they use a customized toolchain that i) reserves a register for instruction counting and ii) instruments the start of each basic block with an instruction to increase the count by the number of instructions in the block.
Although counting instructions is sufficient for their purposes, they acknowledge that counting CPU cycles would be a more accurate metric of CPU effort.

In the \emph{Varys} system, Oleksenko et al.~\cite{oleksenko_varys:_2018} use a rough count of instructions to estimate the frequency of AEX events, since some side-channel attacks require an unusually high frequency of AEX events.
Again, counting instructions is sufficiently accurate for this purpose.
Similarly to our approach, when they require fine-grained timing, they spawn an enclave thread and increment a global variable in a tight loop.

\textbf{Other uses of TSX:}
Shih et al.~\cite{shih_t-sgx:_2017} also use TSX to defend against page-fault side-channel attacks.
Specifically, their T-SGX system encapsulates sensitive enclave code in a TSX transaction and leverages the property that errors (e.g.\ page faults) occurring within a transaction are not reported to the underlying OS, thus making known controlled-channel attacks impractical.

Gruss et al.~\cite{gruss_strong_2017} use TSX to defend against cache side-channel attacks.
Their key insight is that a TSX transaction requires all memory it accesses to remain in the CPU caches for the duration of the transaction.
A premature eviction of some memory results in a transaction abort.
Code that exhibits secret-dependent control flow or data memory accesses is typically vulnerable to cache-base side channel attacks.
By encapsulating such code in TSX transactions, their \emph{Cloak} system ensures that if a transaction completes, all sensitive code and data must have remained in the CPU caches, otherwise the transaction would abort and roll back any memory changes.

One challenge faced by Cloak is that code and data in a transaction's read set are \emph{not} protected against a malicious sibling hyperthread, which can mount cache side-channel attacks from outside the enclave using the L1 and L2 caches it shares with the thread in the enclave.
To enforce that two threads remain in the enclave, they have each thread write a unique marker to each other's SSA, making it part of the read set of the transactions.
An interrupt of either thread overwrites the marker in the SSA, and can be detected from the other thread.
We use this technique to detect interruptions of the worker thread from our timer thread.

\textbf{Securing outsourced computation:}
Rutkowska~\cite{graphene-ng} recently described work towards running arbitrary payloads in SGX enclaves in the context of the Golem network that outsources computation to individuals.
This is complementary to \name{}, and could make use of our mechanisms for accurately and reliably measuring and reporting resource consumption from these small ephemeral service providers.

Similarly, Brenner et al.~\cite{brenner_secure_2017} presented \emph{Vert.x Vault}, a design for running Java-based Vert-x micro services in SGX enclaves, and more broadly, the \emph{SecureCloud} project~\cite{kelbert_securecloud:_2017} aims to secure cloud-based micro services using hardware security mechanisms.
On the client-side, TrustJS~\cite{goltzsche_trustjs:_2017} allows server-supplied JavaScript to be run in an enclave. 
If required, the \name{} resource measurement mechanisms could provide fine-grained resource measurement for any of these services.

These mechanisms can also be used in other SGX-based services, even if they do not support arbitrary computation.
For example, they could form the basis of an auditable billing policy for \emph{SecureKeeper}~\cite{brenner_securekeeper:_2016}, an enhanced version of Apache ZooKeeper that protects data using SGX.

\section{Conclusion \& Future Work}

The emerging Function-as-a-Service (FaaS) paradigm provides numerous benefits, especially in terms of allowing small ephemeral entities (e.g.\ individuals) to offer comparable services to large cloud providers.
However, FaaS significantly exacerbates the challenges of i) ensuring the integrity of outsourced computation, ii) minimizing the information leaked to the service provider, and iii) accurately measuring and reliably reporting computational resource usage.

We introduce \name{}, the first architecture and implementation of FaaS to provide strong security and accountability guarantees backed by Intel SGX. 
We have demonstrated that our design can support existing FaaS workflows by implementing and integrating it with the OpenWhisk FaaS framework.
Our \name{} resource measurement functionality adds less than 6.3\% overhead when applied to the Duktape JavaScript interpreter inside an SGX enclave. 
Our resource measurement mechanisms, based on Intel TSX, can accurately and reliably measure the compute time, memory, and network usage of a function, and can also be used in other applications beyond FaaS.
As future work, we plan to extend \name{} to support functions written in other languages, and to support other types of hardware-based TEEs.

\section*{Acknowledgements}
This work was supported in part by Business Finland (CloSer project, 3881/31/2016) and by Intel (Intel Collaborative Research Institute for Collaborative Autonomous and Resilient Systems, ICRI-CARS).
Andrew Paverd was supported in part by a Fulbright Cyber Security Scholar Award.

\bibliographystyle{IEEEtranS}
\balance
\bibliography{faas-sgx}

\end{document}